\documentclass[conference]{IEEEtran}
\IEEEoverridecommandlockouts
\usepackage{cite}
\usepackage{amsmath,amssymb,amsfonts}
\usepackage{algorithmic}
\usepackage{graphicx}
\usepackage{textcomp}
\usepackage{xcolor}
\usepackage{tabularx}


\def\BibTeX{{\rm B\kern-.05em{\sc i\kern-.025em b}\kern-.08em
    T\kern-.1667em\lower.7ex\hbox{E}\kern-.125emX}}

\begin{document}

\title{Reconstructing ERP Signals \\
Using Generative Adversarial Networks \\
for Mobile Brain-Machine Interface
}

\author{\IEEEauthorblockN{Young-Eun Lee}
\IEEEauthorblockA{\textit{Dept. Brain and Cognitive Engineering} \\
\textit{Korea University}\\
Seoul, Republic of Korea \\
ye\_lee@korea.ac.kr}
\and
\IEEEauthorblockN{Minji Lee}
\IEEEauthorblockA{\textit{Dept. Brain and Cognitive Engineering} \\
\textit{Korea University}\\
Seoul, Republic of Korea \\
minjilee@korea.ac.kr}
\and
\IEEEauthorblockN{Seong-Whan Lee}
\IEEEauthorblockA{\textit{Dept. Artificial Intelligence  } \\
\textit{Korea University}\\
Seoul, Republic of Korea \\
sw.lee@korea.ac.kr}

\thanks{© 20xx IEEE. Personal use of this material is permitted. Permission	from IEEE must be obtained for all other uses, in any current or future media, including reprinting/republishing this material for advertising or promotional purposes, creating new collective works, for resale or redistribution to servers or lists, or reuse of any copyrighted component of this work in other works. }
\thanks{This work was supported by the Institute for Information \& Communications Technology Planning \& Evaluation (IITP) grant funded by the Korea government (No. 2017-0-00451, Development of BCI based Brain and Cognitive Computing Technology for Recognizing User’s Intentions using Deep Learning; No. 2015-0-00185, Development of Intelligent Pattern Recognition Softwares for Ambulatory Brain Computer Interface).}
}
\maketitle

\begin{abstract}
Practical brain-machine interfaces have been widely studied to accurately detect human intentions using brain signals in the real world. However, the electroencephalography (EEG) signals are distorted owing to the artifacts such as walking and head movement, so brain signals may be large in amplitude rather than desired EEG signals. Due to these artifacts, detecting accurately human intention in the mobile environment is challenging. In this paper, we proposed the reconstruction framework based on generative adversarial networks using the event-related potentials (ERP) during walking. We used a pre-trained convolutional encoder to represent latent variables and reconstructed ERP through the generative model which shape similar to the opposite of encoder. Finally, the ERP was classified using the discriminative model to demonstrate the validity of our proposed framework. As a result, the reconstructed signals had important components such as N200 and P300 similar to ERP during standing. The accuracy of reconstructed EEG was similar to raw noisy EEG signals during walking. The signal-to-noise ratio of reconstructed EEG was significantly increased as 1.3. The loss of the generative model was 0.6301, which is comparatively low, which means training generative model had high performance. The reconstructed ERP consequentially showed an improvement in classification performance during walking through the effects of noise reduction. The proposed framework could help recognize human intention based on the brain-machine interface even in the mobile environment.\\
\end{abstract}

\begin{IEEEkeywords}
brain-machine interface, ambulatory BMI, mobile BMI, generative adversarial networks, event-related potentials
\end{IEEEkeywords}

\begin{figure*}[t]
	\begin{center}
		\includegraphics{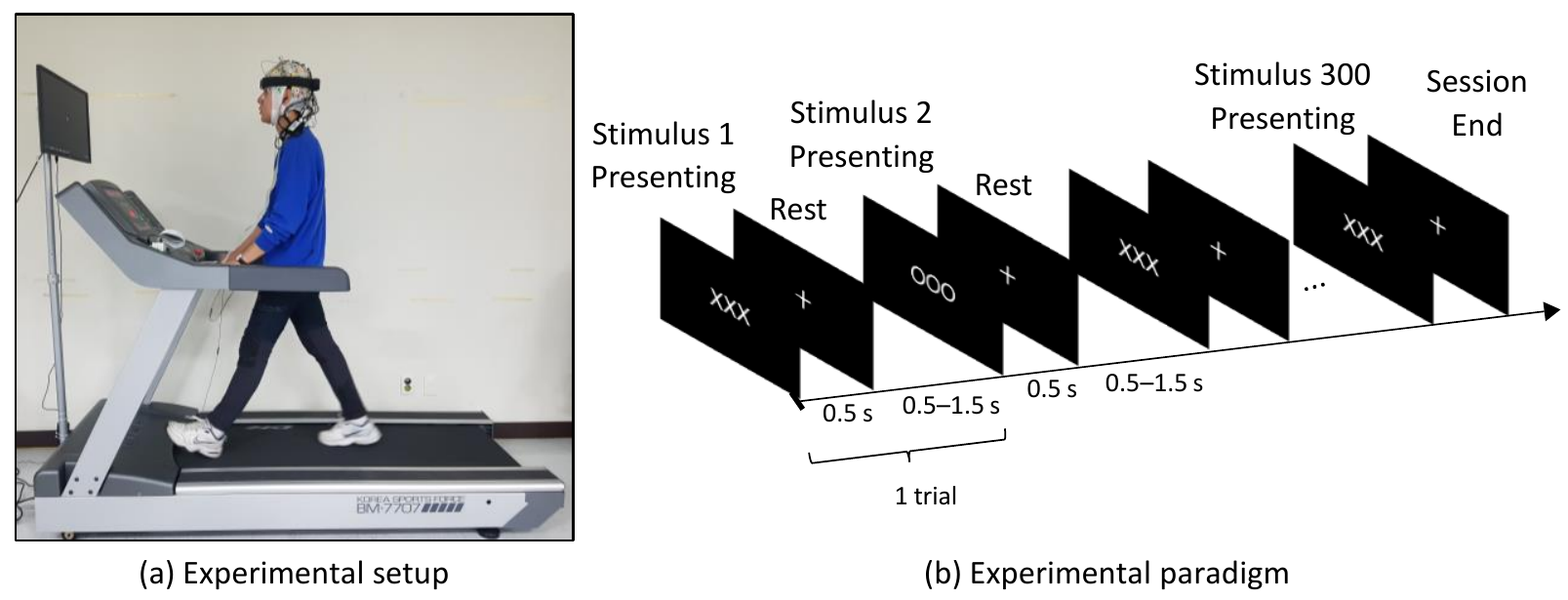}
		\caption{Experimental design. (a) Experimental setup. All subjects are asked to stand or walk on the treadmill in front of a display. (b) Experimental paradigm. ERP paradigm including `target' and `non-target' is used in this experiment.}
		\label{fig1}
    \end{center}
\end{figure*}  

\section{Introduction}
% Ambulatory BMI 문제점 및 현주소
Brain-machine interfaces (BMIs) are technical systems that enable impaired people to communicate and control machines or robots by decoding human intention from brain signals~\cite{wolpaw2002brain, lee2016motor,zhu2016canonical,ding2013changes}. There are many state-of-the-art BMI systems to increase the performance of identifying user intention in a laboratory condition~\cite{lee2015subject,chen2016high}. In particular, BMIs under an ambulatory condition are important issues for practical BMIs to recognize human intention in the real world~\cite{gramann2010visual,luu2017real,malcolm2019long}. However, the movement artifacts can have difficulty detecting user intention because they affect electroencephalography (EEG) signals with large magnitudes. These artifacts could arise from head movement, electromyography, muscle activity, skin, and cable movement~\cite{symeonidou2018effects}. Several studies about BMIs in the ambulatory environment have been actively conducted by applying artifact removal methods in the pre-processing phase~\cite{nordin2018dual,blum2019riemannian} or using the high-tech methodology in the feature extraction or classification phase to better understand user intention~\cite{kwak2015lower, Lee2020decoding}. 
These processes to reduce the effects of artifacts are essential for practical BMIs.

% GAN
Generative models produced the data distribution through decoding progress commonly applying for audio, images, or videos. While the two trainable models contested each other, they learned in the direction that they do not know whether data are generated or not. Recently, a novel generative model is introduced using deep neural networks to represent and reconstruct such as generative adversarial networks (GANs)~\cite{goodfellow2014generative} and many of its advanced version. GANs are machine learning frameworks consisting of two neural networks that contest each other in a zero-sum game. 
Deep convolutional GANs (DCGANs)~\cite{radford2015unsupervised} are an advanced model of GANs which can train models with convolutional layers to be more stable than normal GANs. Auxiliary classifier GANs (ACGANs)~\cite{odena2017conditional} are also one of the improved version of GANs, which can train the class information of data at the same time to improve the generative data. Various advanced versions of GANs are used differently depending on purposes such as data augmentation and style of image changes.

% EEG데이터에서 DEEP learning 및 GAN 사용
Recently, many studies have been reported to improve classification performance by using deep neural networks in EEG data~\cite{kwak2017convolutional, lee2018spatio}. However, researchers have struggled to use traditional deep neural networks since EEG signals have different characteristics from typical input of deep neural networks. EEG has dynamic time series data and the amplitude of artifact is higher than the sources which contain the human intention. 
Thus, there are several attempts to fit EEG signals into deep neural networks. In Schirrmeister et al.\cite{schirrmeister2017deep}, they introduced deep ConvNets which is EEG-fitted convolutional neural networks (CNN) and compared with traditional classifier for motor imagery, having much higher performance. 
EEGNET~\cite{lawhern2018eegnet} was developed for EEG signals during BMI paradigms including P300 visual-evoked potential (VEP) paradigm.
Moreover, a few papers recently used GANs in EEG data to generate another EEG data. In Hartmann et al.\cite{hartmann2018eeg}, they generated EEG signals of hand movement using GANs with different architectures, showing the signals generated well in time-series and frequency spectra. In addition, GANs were trained to classify and generate EEG data for driving fatigue~\cite{panwar2019semi}. To date, most studies applying GANs to EEG data are used for data augmentation purposes to improve classification performance.

% GAN 을 noise reduction 으로 사용하기 위한 연구
GANs are used for noise reduction in a few studies. In Wolterink et al.\cite{wolterink2017generative}, they reduced noises in computed tomography (CT) data using GANs with convolutional neural networks to minimize voxelwise loss. As a result, they produced more accurate CT images. Another researcher~\cite{choi2019cycle} performed the reduced noise in the CT image using models inspired by cycle-GANs~\cite{zhu2017unpaired} and PatchGan~\cite{isola2017image}. These studies demonstrated that not only normal image and audio data but also brain-related data such as brain imaging are applied for noise reduction using GANs.

% In this paper
In this paper, we proposed the reconstruction framework of event-related potential (ERP) from noisy EEG signals during walking. To reconstruct the EEG signals, we utilized a generative model framework inspired by EEGNET~\cite{lawhern2018eegnet}, DCGANs~\cite{radford2015unsupervised}, and ACGANs~\cite{odena2017conditional} and then classified the ERP signals using convolutional discriminative models. To make the latent variables for the generative model, we used the pre-trained model consisted of convolutional neural networks, encoding noisy EEG signals in the ambulatory environment.
We hypothesized that reconstructed EEG would contain ERP components but not have artifacts. We performed subject-dependent and subject-independent training sessions. We also evaluated the reconstructed ERP with the visual inspection, ERP performance, and the loss of the generative model. This work could be a noise reduction method and extracting user intention methods.

\begin{figure*}[t]
	\begin{center}
		\includegraphics{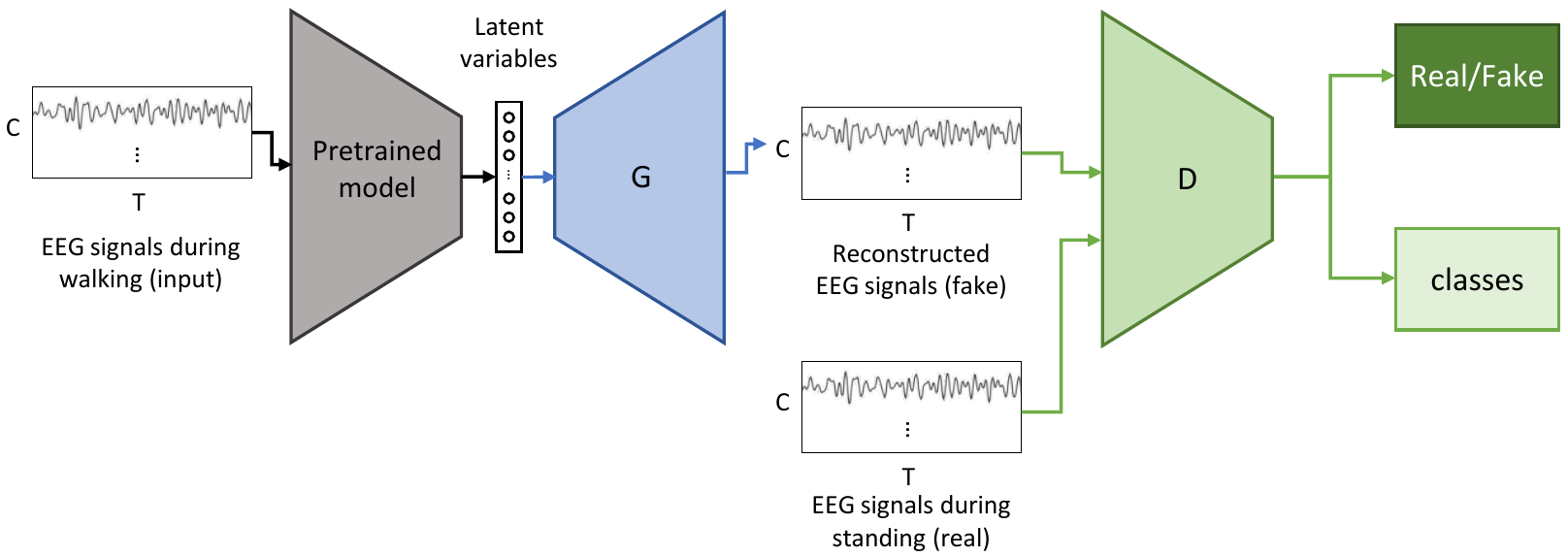}
		\caption{Proposed GAN architectures for ERP reconstruction with the input of EEG signals. C indicates the number of channels and T indicates the time samples in a segment. The input of EEG signals during walking goes through the pre-trained model and latent variables are produced. Latent variables are the input of the generative model $G$ to reconstruct ERP signals. Discriminative model $D$ distinguishes whether the input signals are fake (reconstructed signals) or real (EEG signals during standing) and the classes.}
		\label{fig2}
    \end{center}
\end{figure*}

\section{Materials and Methods}
\subsection{Experimental Setup}
\subsubsection{subjects}
Eighteen healthy young subjects (four females, age 24.5 $\pm$ 3.1 years) were included in this experiment. None of the subjects had a history of neurological, psychiatric, or any other pertinent disease that otherwise might have affected the experimental results. All subjects gave their written informed consent before the experiments. All experiments were carried out corresponding to the Declaration of Helsinki. This study was reviewed and approved by the Korea University Institutional Review Board (KUIRB-2019-0194-01). 

The subjects were on the treadmill at 80 ($\pm$5) cm in front of a 24 inch LCD monitor (refresh rate: 60 Hz, resolution: 1920 $\times$ 1080) and stood, walked at 1.6\,m/s during the BMI paradigms (Fig.~\ref{fig1}-(a)).

\subsubsection{Data acquisition}
We used a wireless interface (MOVE system, Brain Product GmbH) and Ag/AgCl electrodes to acquire EEG signals from the scalp and Smarting System (mBrainTrain LLC) to record EEG signals. The cap electrodes were placed according to the 10-20 international system at locations in 32 channels: Fp1, Fp2, AFz, F7, F3, Fz, F4, F8, FC5, FC1, FC2, FC6, C3, Cz, C4, CP5, CP1, CP2, CP6, P7, P3, Pz, P4, P8, PO7, PO3, POz, PO4, PO8, O1, Oz, and O2. The impedance was maintained below 10 $k\Omega$. We set the sampling rate at 500 Hz.

\subsubsection{Paradigm}
We acquired data of ERP based on the OpenBMI (http://openbmi.org)~\cite{lee2019eeg}, BBCI (http://github.com/bbci/bbci\_public)~\cite{krepki2007berlin}, and Psychophysics (http://psychtoolbox.org) toolboxes~\cite{kleiner2007what} in Matlab (The Mathworks, Natick, MA).

ERP is an electrical potential induced in the central and parietal cortex in response to particular cognitive tasks~\cite{lee2018high}. Attention on target induces ERP components such as N200 and P300 which have task-relevant negative peaks 200\,ms and positive peaks 300\,ms after a target stimulus. In this experiment, this paradigm is executed with the target (`OOO') and non-target (`XXX') characters. The ratio of the target was 0.2 and the number of total trials is 300. In a trial, a stimulus was presented for 0.5\,s, and showing cross to take a rest for randomly 0.5--1.5\,s (Fig.~\ref{fig1}-(b)).
  
\subsection{Generative Adversarial Networks}
GANs were a novel generative model, consisting of two adversarial models which are generative model and discriminative model. The generative model produced the data which is the input of the discriminative model which distinguished whether the data is reconstructed or standing data and its class. The generative model is trained by mapping from latent variables to purpose data distribution. While the latent variables are normally random noise, we put feature vectors through the pre-trained model including CNN architectures from every single noisy EEG signal during walking. The discriminative model learned to distinguish whether EEG signals during or reconstructed fake signals (validity) and classify their classes. The contest progressed well-made fake signals versus precise discriminator. Thus, the purpose of the generative model is to increase the validity loss of the discriminative model which tried to reduce its loss.

Fig.~\ref{fig2} showed the frameworks of used GANs with the input of EEG signals to detect ERP signals. We used the pre-trained CNN model for the first encoder from the input. The generator $G$ reconstructed ERP signals from the input of noisy EEG signals. The discriminator $D$ distinguished the reconstructed data and data during standing whether real or fake and the classes. 

\begin{table}[t]
\caption{Generative Architecture Layer and Output Shape}
\begin{center}
\begin{tabular}{|c|c|c|c|}
\hline
\multicolumn{2}{|c|}{\textbf{Generator}}&\multicolumn{2}{|c|}{\textbf{Discriminator}} \\
\cline{1-4} 
\textbf{\textit{Layers}}& \textbf{\textit{Output}}& \textbf{\textit{Layers}}& \textbf{\textit{Output}} \\
\hline
Dense & T/2 &  Conv2D & C$\times$T$\times$8 \\
Reshape &2$\times$T/16$\times$4& ReLU &C$\times$T$\times$8 \\
Batch normalization &2$\times$T/16$\times$4 &dropout& C$\times$T$\times$8 \\
up sampling &4$\times$T/4$\times$4 & permute &8$\times$T$\times$C \\
zero padding &5$\times$T/4$\times$4 & conv2D &8$\times$T$\times$8 \\
conv2D &5$\times$T/4$\times$8& ReLU & 8$\times$T$\times$8 \\
ReLU &5$\times$T/4$\times$8 & max pooling &4$\times$T/4$\times$8\\
Batch normalization &5$\times$T/4$\times$8& dropout &4$\times$T/4$\times$8\\
up sampling &10$\times$T$\times$8& Batch normalization &4$\times$T/4$\times$8\\
conv2D &10$\times$T$\times$C&conv2D&4$\times$T/4$\times$4\\
ReLU &10$\times$T$\times$C& ReLU &4$\times$T/4$\times$4\\
Batch normalization &10$\times$T$\times$C & max pooling &2$\times$T/16$\times$4\\
up sampling &20$\times$T$\times$C& dropout &2$\times$T/16$\times$4\\
permute & C$\times$T$\times$20& Batch normalization &2$\times$T/16$\times$4\\
conv2D & C$\times$T$\times$1& flatten & T/2 \\
ReLU & C$\times$T$\times$1& & \\

\hline
\end{tabular}
\label{tab1}
\end{center}
\end{table}

\begin{figure*}[t]
	\begin{center}
		\includegraphics{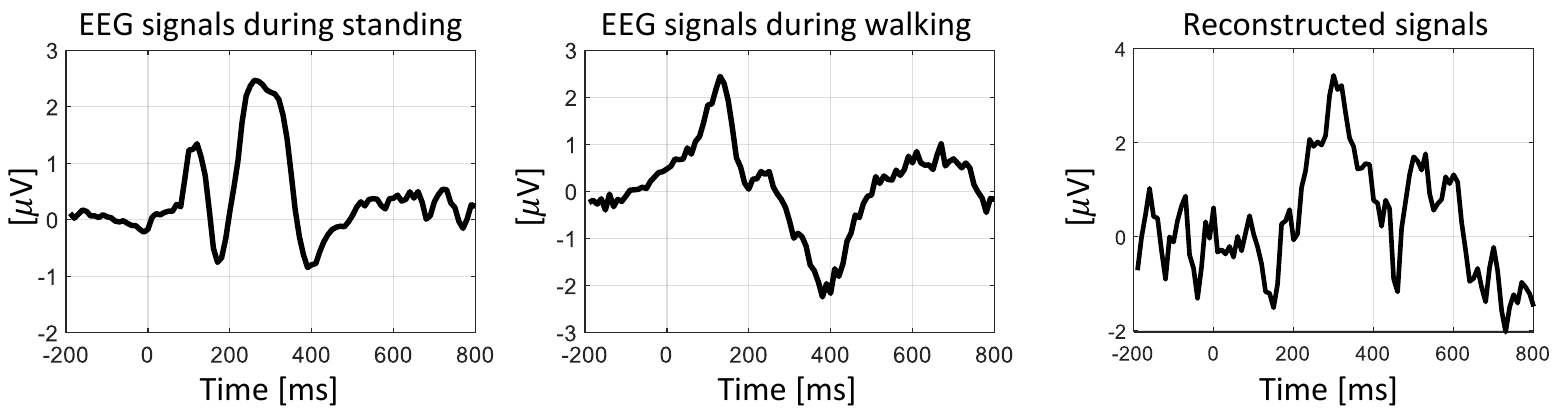}
		\caption{Grand average of EEG signals during standing and walking and reconstructed EEG signals. The reconstructed signals are derived from the raw EEG signals during walking through the proposed framework.}
		\label{fig3}
    \end{center}
\end{figure*}

\subsubsection{Discriminative model}
The discriminative model was inspired by EEGNET~\cite{lawhern2018eegnet} fitted to EEG data consisting of convolutional layers, leaky rectified linear unit (ReLU), dropout, max pooling, and batch normalization. While the input of the discriminative model is cleaned EEG signals and reconstructed EEG signals, the output is whether real or fake and the classes. 

\subsubsection{Generative model}
The generative model was similar to the opposite of the structure of the discriminative model, which consisting of dense, convolutional layer, batch normalization, up sampling, and zero padding. The latent variable was produced by convolutional pre-trained model inputting of each noisy EEG samples. While the input of generative model is latent variables, the output is the reconstructed signals which are the same shape of noisy EEG samples.

\subsubsection{Training}
Generative model $G$ generated the data distribution from latent values and discriminative model $D$ distinguished the data whether fake or not (validity). 
We used an advanced model of GANs that can offer not only the validity of data but also classification. The $D$ of GANs is trained to maximize the log-likelihood of validity $L_S$ and classification $L_C$, which can be represented by following as:

\begin{align}
    L_S = E[\log{P}(S=&real|X_{real})]  + \\ \nonumber
    & E[\log{P}(S=fake|X_{fake})]
\end{align}
\begin{align}
    L_C = E[\log{P}(C=&c|X_{real})] + \\ \nonumber
    & E[\log{P}(C=c|X_{fake})]
\end{align}
where $D$ is trained to maximize $L_S + L_C$, $G$ is trained to maximize $L_S - L_C$.

All data were applied high-pass filter at 0.5 Hz and epoched into the time-interval between --200\,ms and 800\,ms based on the trigger. In reconstructed EEG signals, we selected one channel at Pz where the amplitude of ERP is apparent. To divide the dataset, we performed cross-validation for each subject which we left one subject as test set and trained rest subject dataset. This is named leave-one-subject-out which is cross validation inter trials but inter subjects~\cite{fahimi2019inter}. For the pre-trained data, we trained neural networks with walking dataset and tested standing dataset. As we managed the imbalanced data for training, we reduced the number of non-target trials as same as target trials. We trained GANs in each batch, the size with 32, using loss function cross-entropy for validity and classification of ERP, and mean squared error for generative model loss of calculating between the original input signals and reconstructing signals. We trained model using Adam optimizer~\cite{kingma2014adam} with a learning rate of 0.0002 and the exponential decay rates of the 1st moment estimates of the gradient 0.5~\cite{radford2015unsupervised}. 

\begin{figure}[t]
	\begin{center}
       % \vspace*{\floatsep}
		\includegraphics{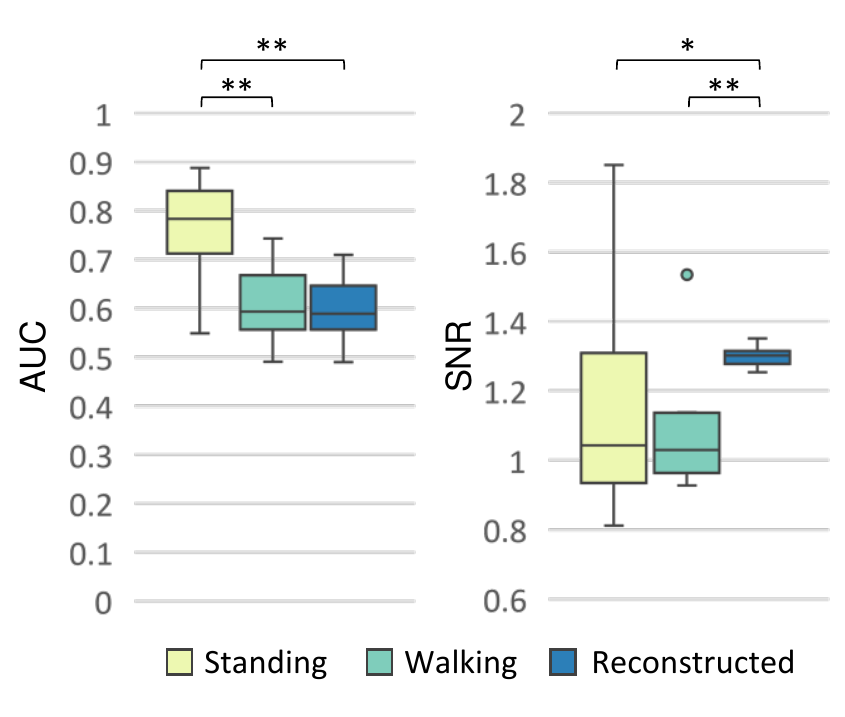}
		\caption{The performance in subject-independent session. Standing and walking refer to raw EEG signals during standing and walking, respectively. The restructured refers to the regenerated signals through the proposed framework based on EEG signals during walking. One asterisk and two asterisks indicate $5\,\%$ and $1\,\%$, significance level, respectively.}
		\label{fig4}
    \end{center}
\end{figure}  

\begin{figure*}[t]
	\begin{center}
       % \vspace*{\floatsep}
		\includegraphics{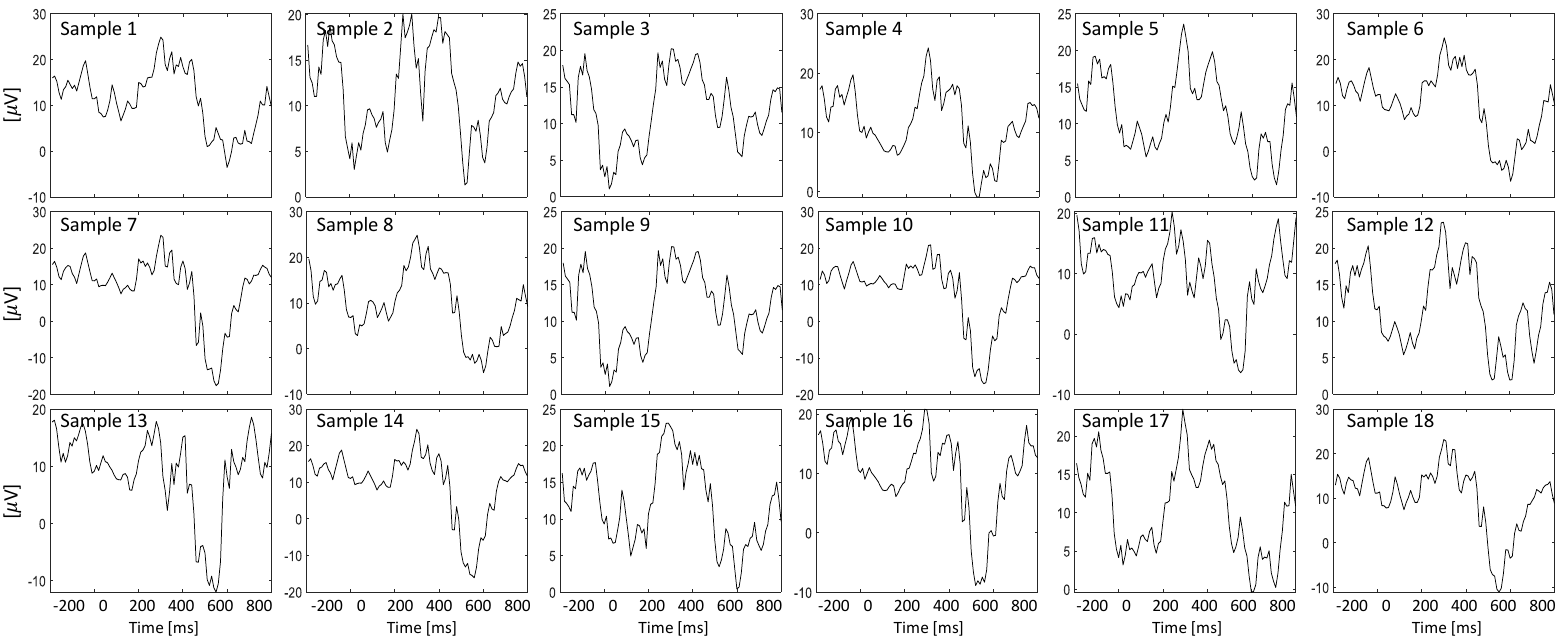}
		\caption{Samples of reconstructed EEG signals. These signals were generated through the proposed framework based on EEG signals during walking. Each sample showed the time-series plot between --200\,ms and 800\,ms.}
		\label{fig5}
    \end{center}
\end{figure*} 

\section{Results and Discussion}
We performed the training in subject-independent sessions. We have presented quantitative metrics indicating the training quality and diversity. Generative model and discriminative model were evaluated by visual inspection, classification performance, and signal-to-noise ratio (SNR). The statistic analysis of a two-tailed paired t-test was performed in the confidence level of 95\% and 99\%.

\subsection{Discriminative Model}

\subsubsection{Visual inspection}
% visual inspection
Fig.~\ref{fig3} showed the grand average of noisy EEG signals during standing, EEG signals during walking, and reconstructed EEG signals from noisy signals. Although the EEG signals during standing had apparent N200 and P300 components, the signals during walking in the mobile environment had the lower amplitude at N200 and P300 because of the great amplitude of artifacts. The reconstructed EEG data also had strong N200 and P300 components in signals as standing signals. The figure showed the generators made data similar to standing data having high SNR.

% classification accuracy
\subsubsection{Classification performance}
Fig.~\ref{fig4} showed the accuracy of standing data, noisy data during walking at 1.6\,m/s, and reconstructed data. The classification performance was calculated by areas under the receiver operating characteristic curve (AUC). The AUC of EEG signals during walking decreased rather than standing condition because of the distortion of EEG signals with great amplitude in the mobile environment. However, the averaged AUC of reconstructed data was similar to walking condition.

% SNR
\subsubsection{Signal quality}
The SNR is an essential indicator of noise reduction, which is calculated by the root mean square (RMS) of the amplitude of peaks at P300 divided by RMS of the average amplitude of pre-stimulus baseline (-200\,ms -- 0\,ms) at channel Pz~\cite{schimmel1967reference}. Fig.~\ref{fig4} showed the average of SNR for standing, walking, and reconstructed data. The SNR during walking decreased than standing data because of data distorting due to motion artifacts. However, applying GANs led to an increase in SNR significantly ($p<0.01$). This is because the amplitude of noise at pre-stimulus would be reduced a lot rather than P300 in reconstructed EEG signals. The SNR of reconstructed signals is even higher than EEG signals during standing.

\subsection{Generative Model}
Fig.~\ref{fig5} showed the samples of reconstructed EEG signals. As can be seen, each sample had N200 and P300 components, which means that the generative model tried to reconstruct the signals containing ERP-related components. Moreover, the signals also similar to raw signals during walking which were original signals of their signals, so that they can retain the information of origins.
A few samples seem to still contain artifacts in pre-stimulus time comparing to the EEG signals during standing.
Fig.~\ref{fig5} also demonstrated the diversity of reconstructed EEG signals. Each sample of reconstructed EEG signals had different shapes, which means the generative model was trained with diversity, and each input of noisy signals impact the output. 
% MSE
The MSE loss of the generative model was 0.6301, which is quite low value comparing to other training loss. 

\section{Conclusion}
In this paper, we proposed the ERP reconstruction framework from noisy signals during walking using GANs. In the mobile environment such as walking or running, the EEG signals contain great amplitude of artifacts, which is much huger than user intention. Therefore, there have been many challenges to enhance the BMI performance in the mobile environment by developing artifact removal methods, extracting critical features from brain signals, and developing classifiers using deep neural networks. Because of the properties of EEG, the neural networks for EEG would be different from the networks for images. One of the deep neural networks for EEG was deep ConvNets, which considered kernel size regarding channels and time series. Generating EEG signals using the generative model is a novel approach and combining deep ConvNets and generative model provided EEG signals generation. We constructed GANs able to not only generate the data and classify validity but also can do classification. For the dataset, we collected data of EEG signals during walking at 1.6\,m/s on the treadmill, and performed oddball ERP paradigm. 

The reconstructed data through GANs had significant components such as N200 and P300 similar to EEG signals during standing. Moreover, the SNR of reconstructed data was much higher than noisy signals in the walking environment. The reconstructing progress was diverse so that it produced various samples of ERPs containing N200 and P300. The loss of the generative model was comparatively low that means training generator had high performance. Therefore, it is thought to extract or reconstruct ERP components from noisy EEG signals which can enhance the ERP performance. The proposed framework would directly help BMIs in the mobile environment to reduct noise in terms of SNR.
In the future, we would advance the model to improve the classification performance. Moreover, we will train all 32 channels of EEG signals to reconstruct ERP signals.

% references section
\bibliographystyle{IEEEtran}
\bibliography{bibliography}

\end{document}